\RequirePackage[l2tabu, orthodox]{nag}
\documentclass[aps,pra,reprint,superscriptaddress,floatfix,longbibliography]{revtex4-1}

\usepackage{textcomp}
\usepackage{amsmath}
\usepackage{graphicx}
\usepackage{hyperref}
\usepackage{epspdfconversion}
\usepackage{bibentry}

\usepackage{color}
\usepackage{soul} 

\usepackage[export]{adjustbox}

\usepackage{siunitx}
\usepackage{textcomp}

\usepackage{lipsum}

\newcommand{\us}{\SI{}{\micro\second}}
\newcommand{\um}{\SI{}{\micro\m}}
\newcommand{\uev}{\SI{}{\micro\electronvolt}}

\newcommand{\paperexlifetime}{7.7(4)}
\newcommand{\paperTDM}{1.96(8)} 
\newcommand{\paperzpleff}{16(1)} 
\newcommand{\paperradeff}{0.80(9)} 
\newcommand{\papermaxTone}{4.6(1.5)} 
\newcommand{\papertautotal}{0.90(7)}
\newcommand{\papertipconc}{5.2(4)$~\times~$10$^{14}$}

\newcommand{\paperhiB}{5.9(8)$~\times~$10$^{14}$}

\newcommand{\paperconv}{6.2(5)$~\times~$10$^{14}$}

\newcommand{\paperAtwo}{w$^{28}$Si:$^{77}$Se:IB}

\newcommand{\paperCone}{w$^{\rm nat}$Si:$^{\rm nat}$Se:HB}

\newcommand{\paperDone}{$^{28}$Si:$^{77}$Se:LB}

\newcommand{\paperEone}{$^{28}$Si:$^{78}$Se:IB}

\begin{document}

\title{Characterization of the Si:Se$^+$ spin-photon interface}

\author{Adam DeAbreu}
\author{Camille Bowness}
\author{Rohan J.~S.~Abraham}
\author{Alzbeta Medvedova}
\author{Kevin J.~Morse}
\affiliation{Department of Physics, Simon Fraser University, Burnaby, British Columbia V5A 1S6, Canada}

\author{Helge Riemann} 
\author{Nikolay V.~Abrosimov}
\affiliation{Leibniz-Institut f\"{u}r Kristallz\"{u}chtung, 12489 Berlin, Germany}

\author{Peter Becker}
\affiliation{Physikalisch-Technische Bundesanstalt (PTB) Braunschweig, 38116 Braunschweig, Germany}

\author{Hans-Joachim Pohl}
\affiliation{VITCON Projectconsult GmbH, 07745 Jena, Germany}

\author{Michael L.~W.~Thewalt}
\author{Stephanie Simmons}
\email[Corresponding author: ]{s.simmons@sfu.ca}
\affiliation{Department of Physics, Simon Fraser University, Burnaby, British Columbia V5A 1S6, Canada}

\renewcommand{\figurename}{Fig.}

\date{\today}

\begin{abstract}
Silicon is the most developed electronic and photonic technological platform 
and hosts some of the highest-performance 
spin 
and photonic qubits 
developed to date. A hybrid quantum technology harnessing an efficient 
spin-photon interface in silicon would unlock considerable potential by 
enabling ultra-long-lived photonic memories, distributed quantum 
networks, microwave to optical photon converters, and spin-based quantum processors, 
all linked using integrated silicon photonics. 
However, the indirect bandgap of silicon makes identification 
of efficient spin-photon interfaces nontrivial. 
Here we build upon the recent identification of chalcogen donors 
 as a promising 
spin-photon interface in silicon. We determined that the spin-dependent optical 
degree of freedom has a transition dipole moment stronger than previously thought 
(here \paperTDM{} Debye), and the T$_1$ spin lifetime in low magnetic fields is 
longer than previously thought ($>$ 4.6(1.5)
hours). We furthermore determined the optical excited state lifetime (\paperexlifetime{}~ns), and therefore the natural radiative efficiency (\paperradeff{}~\%),
and by measuring the phonon sideband, determined the zero-phonon emission 
fraction (\paperzpleff{}~\%).
Taken together, these parameters indicate that an integrated 
quantum optoelectronic platform based upon chalcogen donor qubits in silicon is well
within reach of current capabilities.
\end{abstract}

\maketitle 

\section{Introduction}

A future quantum technology, wherein stored quantum information is communicated over 
a quantum network, will necessarily involve both matter-based qubits and optical photons.
In pursuit of this aim, many potential spin-photon interfaces are being 
actively developed~\cite{Yuan2008, Yilmaz2010, 
Bernien2013}. 
A wide array of defects in semiconductors and insulators 
have attracted attention because of their favourable optical and spin 
characteristics. These include quantum dots in III-V heterostructures~\cite{DeGreve2012}, 
nitrogen-vacancy~\cite{Robledo2011}
and silicon-vacancy~\cite{Rose2018}
 centers in diamond, rare-earth ions in insulators such as 
Nd:YSO~\cite{Zhong2015} and Er:YSO~\cite{MZhong2015}, defects in SiC~\cite{Christle2017, Widmann2014}, and even recent work on donors in 
ZnO~\cite{Linpeng2018}. Notable in its absence from this list is silicon, which, when isotopically 
purified to $^{28}$Si, is host to some of the longest-lived and highest-fidelity spin 
qubits studied to date~\cite{Saeedi_39min_lifetime, Wolfowicz2012, Muhonen2015, Dehollain2015}. Silicon offers high performance integrated single photon detectors~\cite{Akhlaghi2015} in addition to an expansive selection of high quality photonic components~\cite{Mekis2011, Hochberg2013} due to decades of fabrication process development. Furthermore, silicon has a strong $\chi^{(3)}$ nonlinearity and large refractive index that enables dense packing of photonic circuitry. Despite the considerable advantages of these two quantum silicon 
platforms, unifying these technologies through an efficient spin-photon interface 
has proven elusive. 

A few paramagnetic centres in silicon possess spin-dependent optical 
transitions, including shallow donor-bound 
excitons~\cite{Thewalt2007_P31_nuclear} and orbital transitions in
rare earth ions, such as 
erbium~\cite{Yin2013}. However, in the aforementioned cases, the defects only weakly couple to light as 
determined by their small optical transition dipole moments. 
Although recent work has demonstrated evanescent coupling of defects with strong 
transition dipole moments in materials placed adjacent to silicon photonic 
structures~\cite{Kim2017},
 the coupling strengths and photon collection efficiencies are 
inherently limited in such designs. 

The ideal silicon spin-photon interface would be a natively-integrated optical 
center which possess a long-lived spin, a high transition dipole moment, and a high radiative efficiency.
In this work we demonstrate that 
singly-ionized deep chalcogen donors in silicon possess
a strong light-matter interaction, suitable for strong coupling
to silicon photonic cavities at the single-spin level. This offers a clear path towards
chalcogen-based integrated
silicon quantum optoelectronics.

The optical characteristics of substitutional chalcogen donors (specifically sulfur, 
selenium, and tellurium) have been studied for 
decades~\cite{Swartz1980, Grimmeiss_properties_of_se_si, Grimmeiss1982_splitting, Janzen1984_hi_res, steger_se_si_spectrum}.
 It was identified 
that the natural distribution of silicon and chalcogen isotopes act as sources 
of static inhomogeneity in the bulk.
Consequently, ultra-sharp optical linewidths, on the order of \uev{} can be achieved~\cite{steger_se_si_spectrum} by 
working with ensembles of individual chalcogen isotopes in isotopically purified $^{28}$Si.
This remarkable uniformity allowed for the hyperfine 
splitting and the electron spin g-value of the 1s:A ground state of singly-ionized 
$^{77}$Se to be directly observed through optical excitation into the first excited 
state, 1s:T$_2$:$\Gamma_7$. Following this, initial electron spin characterization
at X-band microwave frequencies on $^{77}$Se$^+$ demonstrated promising electron spin 
qubit coherence and lifetime characteristics~\cite{LoNardo2015}
 similar to that of the
shallow donors' ultra-long lived electron spins.

The identification of singly-ionized chalcogen donors as a promising spin-photon 
interface in silicon was only made relatively recently~\cite{Morse2017}, and bounds on 
some key spin and optical parameters of $^{77}$Se$^+$ were determined to support this 
proposal. Key parameters in Ref.~\cite{Morse2017}
included a lower bound on the spin T$_1$ 
lifetime ($>$ 6.2(4) minutes) as well as lower bounds on the optical transition 
dipole moment ($>$ 0.77 Debye), optical excited state lifetime ($>$ 5.5 ns), 
as well as an upper bound on the calculated radiative lifetime ($<$ 39 \us). 
In this work we improve upon the bounds on all of these key parameters, including 
the spin T$_1$ time ($>$~\papermaxTone{}~hours) and the transition dipole moment (\paperTDM{} Debye). 
Furthermore, we offer new insights into the optical transition of interest by 
reporting the phonon sideband profile and zero 
phonon line (ZPL) fraction (\paperzpleff{}~\%) and a direct measurement of the excited 
state lifetime (\paperexlifetime{}~ns), and hence the total radiative efficiency (\paperradeff{}~\%).
Lastly, we precisely determine the 
location of the second excited state in the neutral charge state of Se by 
performing Raman spectroscopy. The experimental results presented in Section~\ref{sec:results}
are structured in that order. 

\subsection{The $^{28}$Si:$^{77}$Se$^{+}$ spin-photon system}

Substitutional selenium atoms in silicon are deep double donors. When singly ionized, 
the unpaired spin-$\frac{1}{2}$ electron possesses a hydrogenic orbital structure with a 1s 
ground state. The sixfold degeneracy of the conduction band and the two electron 
spin states give rise to twelve 1s levels which are split by a 
combination of central cell, valley-orbit, and spin-orbit effects. The spin and photon 
degrees of freedom relevant to this work are all contained within these twelve electronic 
1s levels. 

The ground state, 1s:A, possesses two degenerate electron spin levels and has a binding energy of 
$\sim$593~meV~\cite{Grimmeiss1982_splitting}. The first orbital excited states, labeled 1s:T$_2$, are split into
components labeled 1s:T$_2$:$\Gamma_7$ and 1s:T$_2$:$\Gamma_8$, the lower of which, 1s:T$_2$:$\Gamma_7$, possesses
two spin-orbit levels and has a binding energy of $\sim$166~meV~\cite{Grimmeiss1982_splitting}. The
remaining 1s levels, labeled
1s:E, are thought to lie above 1s:T$_2$:$\Gamma_8$,
as is the case for the neutral charge state Se$^0$ and the group V shallow donors, 
but have not been observed in Se$^+$. The optical
transitions between 1s:A and 1s:T$_2$:$\Gamma_7$ are forbidden
according to effective mass theory (EMT) but are symmetry-allowed, and approximately 
427~meV, or 2.9~\um{},~\cite{Grimmeiss1982_splitting}
 which is in the mid-infrared optical band. Further details on the orbital structure 
of this system are given in Ref.~\cite{Morse2017} and references therein. 

Additionally, the $^{77}$Se$^+$ isotope possesses a spin-$\frac{1}{2}$ nuclear spin and a corresponding 
$A$ $\approx$ 1.66 GHz hyperfine~\cite{Morse2017, steger_se_si_spectrum} interaction within the 1s:A electronic manifold. This gives rise 
to a ground state spin Hamiltonian shared by that of the neutral shallow donor $^{31}$P and given by
\begin{equation}
\mathcal{H} =  \frac{g_e \mu_B}{h} B_0 S_z -  \frac{g_n \mu_N}{h} B_0 I_{z} + A \vec{S} \cdot \vec{I},
\end{equation}
where $A$, the hyperfine constant, and $g_e$ (2.0057) and $g_n$ (1.07),
the electron and nuclear $g$-factors respectively, 
are specific to $^{77}$Se$^+$. Here $\mu_B$ and $\mu_N$ are the Bohr and nuclear magnetons, $h$ is the Planck 
constant, and $\vec{S}$ and $\vec{I}$ are the spin operators of the electron and nucleus. At zero 
magnetic field, this spin Hamiltonian results in split 1s:A energy levels defined by electron-nuclear 
spin singlet and triplet states. The 1s:T$_2$:$\Gamma_7$ state possesses no such splitting and therefore 
these levels form a lambda transition~\cite{Grimmeiss1982_splitting} which can be spectrally resolved in the bulk~\cite{steger_se_si_spectrum, Morse2017}.  The allowed 
magnetic resonance transitions from the singlet state $S_0$ to the triplet states $T_-$, $T_0$, $T_+$ support 
long lived qubits~\cite{Morse2017},
particularly across the $S_0 \Leftrightarrow T_0$ transition which is a 
`clock transition'~\cite{Wolfowicz2013} at zero field.

\section{Experimental results}
\label{sec:results}

\subsection{Singlet-triplet T$_1$ temperature dependence}

\begin{figure}[htbp!]
	\centering
	\adjincludegraphics[width = 1\columnwidth, trim = {0 0 0 0}, clip]{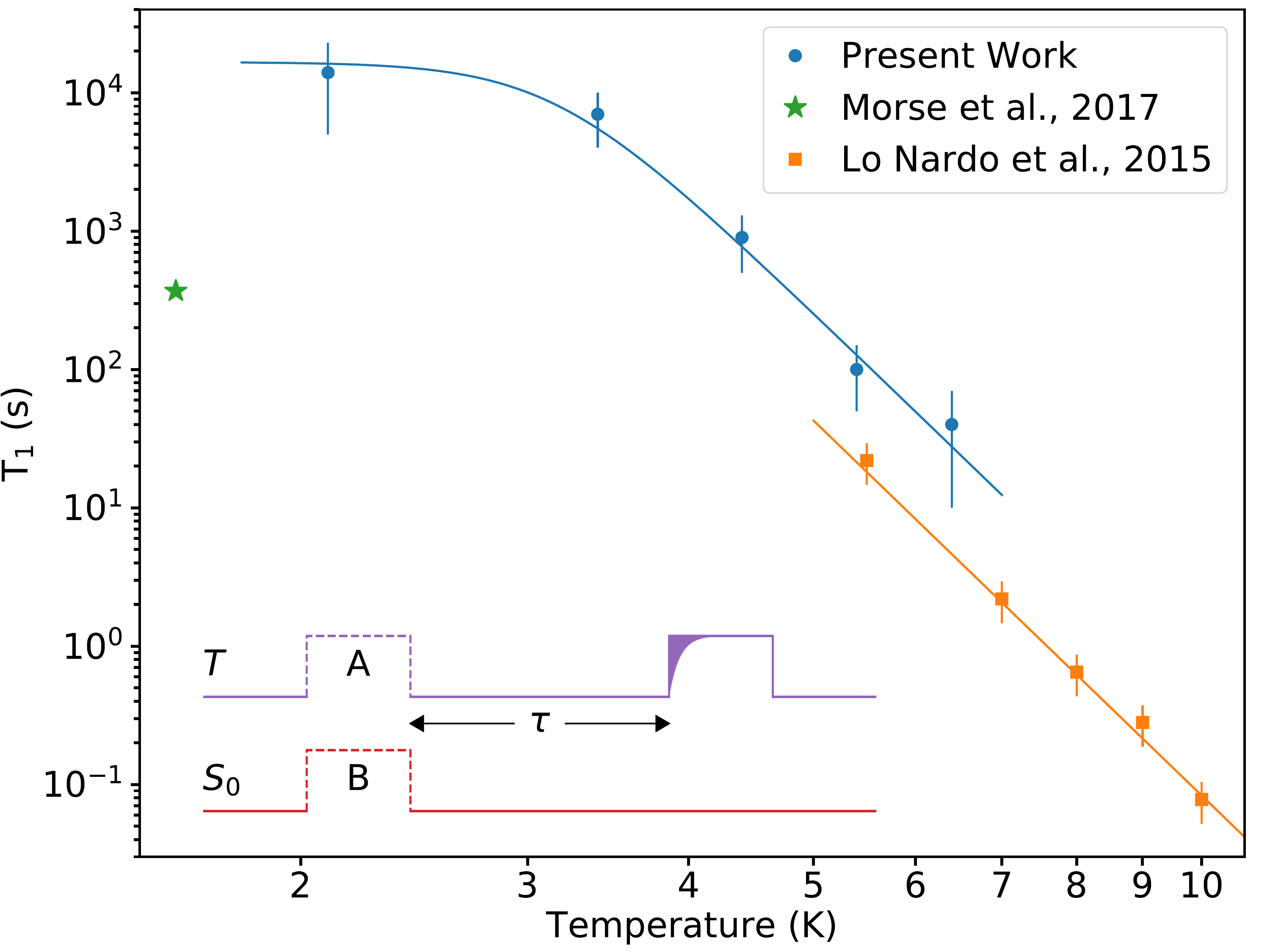}
	\caption{Temperature dependence of $^{77}$Se$^+$ electron-nuclear spin singlet/triplet T$_1$ taken near Earth's magnetic field (blue), revealing a low temperature limit of 
T$_1$~=~4.6(1.5)~hours, and comparison with published data, collected with less blackbody
 shielding (green) reprinted from Ref.~\cite{Morse2017} with permission, as well as electron spin
  T$_1$ data taken at 0.35~T (orange) reprinted from Ref.~\cite{LoNardo2015} with permission. 
  (\textbf{Inset}) Schematic diagram of optical pumping and readout sequence to measure singlet/triplet
   T$_1$. For a given wait time, $\tau$, the total remaining polarization signal is measured 
   as the difference between two integrated absorption transient areas, both measuring the
   population of the triplet state (solid), after two different
   initialization pulses (dashed). First (A) after initializing into the singlet by pumping
   on the 1s:A:$T$ $\Leftrightarrow$ 1s:T$_2$:$\Gamma_7$ transition, here
   labeled $T$, and secondly (B) after initializing into the triplet by pumping 
   on the 1s:A:$S_0$ $\Leftrightarrow$ 1s:T$_2$:$\Gamma_7$ transition, here labelled $S_0$.}
\label{fig:DT1}
\end{figure}

The spin equilibration time constant, T$_1$, of the $^{77}$Se$^+$ singlet/triplet qubit in 
Earth's magnetic field and at low temperatures (1.6~K) was previously found to be 
approximately 6 minutes~\cite{Morse2017}.
Although already quite long, this T$_1$ time is shorter 
than the $\sim$30 minute electron T$_1$ of $^{31}$P measured at 1.6~K and 0.35~T, as well as significantly 
shorter than the projected electron T$_1$ times available to $^{31}$P at 1.6~K in Earth's magnetic 
field~\cite{Feher1959}. Six minutes is substantially longer than previously measured $^{77}$Se$^+$ electron 
T$_1$ times collected at higher temperatures~\cite{LoNardo2015}, but shorter than the $\sim$337 hour 
projected T$_1$ time following a T$^9$ dependence fitted from this higher-temperature data and extrapolated 
to 1.6~K (See Fig~\ref{fig:DT1}). Here we elucidate the decay mechanisms affecting the 
$^{77}$Se$^+$ singlet/triplet 
qubit and determine an experimental regime which gives rise to 
a 276(90) minute (4.6(1.5) hour) T$_1$ time. 

The similarities between the $^{31}$P and $^{77}$Se$^+$ systems imply that a number of known 
$^{31}$P electron T$_1$ decay mechanisms, such as the direct~\cite{Feher1959}, 
Raman~\cite{Castner1963}, and concentration-dependent 
decay mechanisms with concentrations above 10$^{16}$~cm$^{-3}$~\cite{Feher1959},
 can apply to $^{77}$Se$^+$. The significantly 
larger valley-orbit splitting between the ground 1s:A and first excited states 
1s:T$_2$:$\Gamma_7$ of $^{77}$Se$^+$ -- at least seven times
greater than the maximum phonon energy -- implies that the Orbach~\cite{Castner1967}
 decay mechanism known to apply to $^{31}$P is irrelevant to $^{77}$Se$^+$. 
 
An additional decay mechanism is known~\cite{Morse2017} to contribute 
to the T$_1$ decay of $^{77}$Se$^+$: incident room-temperature blackbody radiation possesses sufficient 
energy to ionize both neutral and singly-ionized $^{77}$Se, directly effecting T$_1$ via
time-varying local 
charge configurations. 
Under our experimental conditions,
blackbody radiation generated within the cryogenic apparatus is 
negligible compared to the room temperature incident blackbody radiation optically coupled 
to the sample. Correspondingly, this blackbody T$_1$ decay mechanism is largely independent 
of the sample temperature. In contrast, the direct and Raman decay mechanisms display a 
1/T and 1/T$^9$ temperature dependence, respectively~\cite{Feher1959, Castner1963}. Measurements of the electron 
T$_1$ in a low-concentration ($\sim$2~$\times$~10$^{13}$~cm$^{-3}$) $^{77}$Se$^+$ bulk sample 
(sample \paperDone{}, see Supplementary Materials~\cite{DPMSuppMat}) as a function of temperature 
were used to determine the prevalence of these possible mechanisms.

In order to minimize room temperature blackbody effects, optical bandpass filters centered at 2.9~$\um{}$ as well as  
neutral density filters were mounted within the cryogenic assembly in the optical beam path, and the sample 
was shielded from room temperature blackbody radiation from all other directions. 

The singlet/triplet 
qubit was initialized by resonantly pumping one arm of the lambda transition, for example 
initializing into the singlet state by selectively pumping 1s:A:$T$ $\Leftrightarrow$ 1s:T$_2$:$\Gamma_7$, 
as described in Ref.~\cite{Morse2017}. Following this, a mechanical shutter blocked the resonant light. 
After a chosen delay, the remaining spin hyperpolarisation was measured by recording the absorption transient
of unblocked resonant light (see Fig.~\ref{fig:DT1} inset). For a given delay 
time, two different measurements were taken and the difference between their 
absorption transients constituted the measured signal. The first of these measurements 
hyperpolarised into the singlet state and the second into the triplet state.
Both measurements' absorption transients consistently probed the final triplet population. This subtraction 
method ensured that the signal would necessarily decay to zero in the long delay limit where 
the spins reached equilibrium. The singlet/triplet T$_1$ lifetime at a given temperature was determined by iterating this measurement with variable delay times.

The temperature dependence of T$_1$ over the range 2.1 to 6.4~K is shown in Fig.~\ref{fig:DT1}. 
The data is well fit by 1/T$_1$~=~AT$^9$ + B, with A~=~2.0(3)~$\times~10^{-9}$~s$^{-1}$K$^{-9}$ 
and a temperature independent contribution with a low temperature limit of 
T$_1$~=~4.6(1.5)~hours, representing a T$^9$ Raman process and most likely a residual blackbody-related decay process dominating below 2~K. This T$^9$ Raman process is in agreement with Ref.~\cite{LoNardo2015} taken at 0.35~T, which is fit well by 1/T$_1$~=~CT$^9$ (C~=~$1.2~\times~10^{-8}$~s$^{-1}$K$^{-9}$). The disagreement near 5~K may be due to temperature offsets between these two different experimental setups; alternatively, although the T$^9$ relaxation process is expected to be independent of magnetic field for electron spins~\cite{Castner1963}, this may not apply when comparing between singlet/triplet spin qubits and nearly pure electron spin qubits. These trends indicate that a spin T$_1$ of 19~$\pm$~3 ~minutes is available at the easily accessible temperature of 4.2~K.

\subsection{Absorption}
\label{sec:Abs}

In this section we present measurements based on optical absorption spectra. We improve upon previous transition dipole moment estimates, and use this data to provide a concentration conversion factor.

\subsubsection{Transition dipole moment}
\label{sec:TDPM}

The optical interaction strength of a spin-photon interface is characterized by its transition dipole moment, $\mu$. 
The dipole moment can be calculated from absorption spectra of a bulk sample, combined with an accurate defect concentration value, and a known optical path length~\cite{DPMSuppMat}. Previous work~\cite{Morse2017} employed a bulk sample with non-uniform $^{77}$Se$^+$ concentration and consequently only lower bounds on the transition dipole moment could be made.

Here we calculate the transition dipole moment using a selenium diffused, $^{28}$Si:$^{77}$Se, wafer sample
(sample~\paperAtwo{}, see Supplementary Materials~\cite{DPMSuppMat}). 
An absorption spectrum was measured using a Bruker IFS 125HR Fourier transform infrared (FTIR) spectrometer with
gold mirrors, a KBr beamsplitter, and a mercury-cadmium-telluride (MCT) detector to obtain an absorption coefficient spectrum
of the Se$^+$ 1s:A~$\Rightarrow$~1s:T$_2$:$\Gamma_7$ transition. Where the absorption coefficient spectrum is calculated according to:
\begin{equation}
\alpha = \frac{-1}{L}{\rm{ln}}\left(\frac{I_s}{I_0}\right),
\end{equation}
where $I_s$ and $I_0$ are FTIR spectra with and without the sample in the beam path, respectively, 
and $L$ is the length of the sample.

This sample was confirmed to have a near-uniform [Se$^+$] concentration by observing the complete compensation of all boron in the sample~\cite{DPMSuppMat}. In this case one might expect [B] = [Se$^+$] throughout the sample, however, the precise distribution of donors and acceptors in the sample may modify this value. To measure [Se$^+$] precisely, we applied a tip-angle measurement~\cite{Wolfowicz2012}, whose details have been described in Ref ~\cite{Morse2017}. We measured [Se$^+$] = \papertipconc{}~cm$^{-3}$ which is less than the measured [B] of \paperhiB{}~cm$^{-3}$, likely indicating the presence of doubly ionized selenium, Se$^{2+}$, or ionized selenium pairs, Se$_2^{+/2+}$. Combining
this with the absorption coefficient spectrum we calculate a transition dipole moment of $\mu$ = \paperTDM{} Debye~\cite{DPMSuppMat}. This value is more than a factor of 2 higher than the previously established lower bound.

\subsubsection{Selenium conversion factor}
\label{sec:conv}

From the tip-angle concentration and absorption coefficient spectrum we determined a conversion factor, 
\begin{equation}
f = \frac{[\text{Se}^+]}{\int\alpha\,d\nu} = \text{\paperconv{}}~\text{cm}^{-1},
\end{equation}
for the 1s:A~$\Rightarrow$~1s:T$_2$:$\Gamma_7$ zero phonon spectral line, where $\int\alpha\,d\nu$ is the integrated absorption coefficient
spectra of the zero phonon spectral line. Peak conversion factors, $k_{\rm Se+}$~=~[Se$^+$]/$\alpha_{\rm max}$,
are tabulated in the Supplementary Materials~\cite{DPMSuppMat}.

\subsection{Photoluminescence}

The radiative properties -- both the radiative efficiency and the zero phonon line fraction -- of the Se$^+$ spin-photon interface have not been previously established. In this section we report the observation of the phonon-assisted luminescence sideband of the 1s:T$_2$:$\Gamma_7$ $\Rightarrow$ 1s:A optical transition, which reveals a zero phonon line fraction of 16(1)~$\%$. Subsequently we measured the excited state lifetime (\paperexlifetime{}~ns) and compared this with the calculated radiative lifetime to infer a radiative efficiency of \paperradeff{}~$\%$.

\subsubsection{Zero phonon line fraction}
\label{sec:pho}

\begin{figure}[h]
	\centering
	\adjincludegraphics[width = 1\columnwidth, trim = {0.7cm 0 0.7cm 0}, clip]{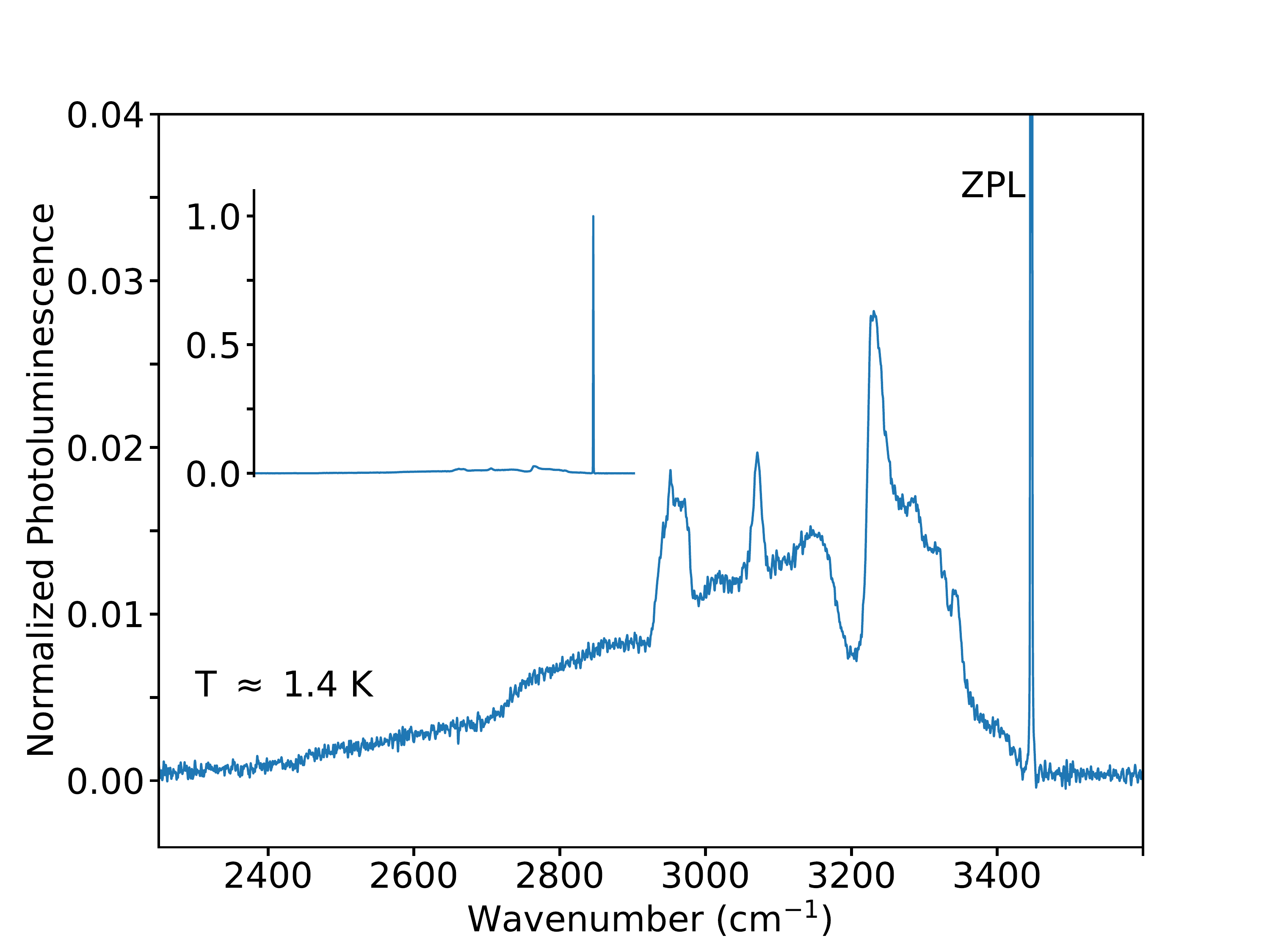}
	\caption{The photoluminescence spectra of the Se$^+$ 1s:T$_2$:$\Gamma_7$ $\Rightarrow$ 1s:A transition taken with a spectral resolution of 1 cm$^{-1}$. (\textbf{Inset}) The normalized spectra showing the relative height of the zero phonon line (ZPL) to the phonon sideband. (\textbf{Main panel}) The same spectra, clipped vertically to display the phonon sideband features. The area of the phonon sideband is 5.6 times larger than the area of the ZPL, revealing a reabsorption-corrected ZPL fraction of \paperzpleff{}~\%.}
	\label{fig:PL}
\end{figure}

Photoluminescence spectra were obtained using a Bruker IFS 125HR FTIR spectrometer with gold mirrors, a CaF$_2$ beamsplitter, and a liquid nitrogen-cooled InSb detector with a 2440 nm long pass filter. A high [Se$^+$] sample (\paperCone{}, see Supplementary Materials~\cite{DPMSuppMat})
 was pumped with 1 W of laser light resonant with the Se$^+$ 1s:A $\Rightarrow$ 2p$_\pm$ transition (4578 cm$^{-1}$, or 2184 ~nm), which was generated using a Cr$^{2+}$:ZnS/Se narrowband tunable laser pumped by an erbium fiber laser (IPG Photonics) operating at 1567 nm. From the excited state 2p$_\pm$, the electron can decay via phonon cascade to 1s:T$_2$:$\Gamma_7$ followed by photon emission to 1s:A. 

The resulting photoluminescence spectrum, including the phonon-assisted sideband, is seen in Fig.~\ref{fig:PL}. The integrated phonon sideband is 5.6 times larger than the area of the zero phonon line, resulting in a ZPL fraction lower bound of 15~\%. After correcting for re-absorption of light given the known ZPL transition dipole moment, which will disproportionately affect the integrated area of the ZPL, we obtain a ZPL fraction of 16(1)~\%.

The total radiative lifetime includes both the zero-phonon and the phonon-assisted radiative pathways, resulting in a total calculated radiative lifetime of $\tau$ = \papertautotal{}~\us{}~\cite{DPMSuppMat}.

\subsubsection{Excited state lifetime and radiative efficiency}
\label{sec:lifetime}

The decay of the 1s:T$_2$:$\Gamma_7$ valley state to the ground state 1s:A can occur through purely radiative, phonon-assisted radiative, and fully nonradiative pathways. The ratio of the  measured 1s:T$_2$:$\Gamma_7$ excited state lifetime to the calculated radiative lifetime reveals the technologically consequential radiative efficiency of this spin-photon interface.

Conventional methods of directly measuring a total luminescence lifetime employ optical pulses and time-resolved,
high-sensitivity detectors which are at least comparable in speed with the transition lifetime of interest. Such sources and detectors are not yet routinely available in the 2.9~\um{} region. Hence, previous to this work, only lower bounds on the total lifetime of this centre were known. Hole-burning measurements, limited by FTIR spectrometer resolution, indicated~\cite{Morse2017} a total excited state lifetime longer than 5.5~ns corresponding to a homogeneous linewidth smaller than 0.001~cm$^{-1}$.

\begin{figure}[h]
	\centering
	\adjincludegraphics[width = 1\columnwidth, clip]{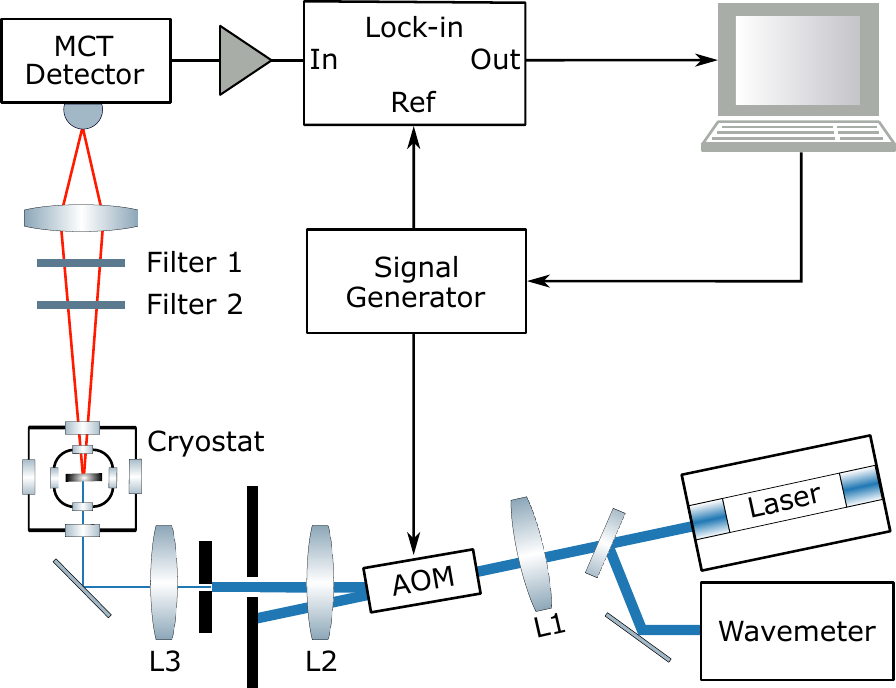}
	\caption{Schematic of experimental set-up. A laser, tuned to the Se$^+$ 1s:A $\Rightarrow$ 2p$_\pm$ transition (4578 cm$^{-1}$, or 2184~nm), whose wavelength was monitored using a pick-off beam routed into a wavemeter, was focused through a lens (L1) to minimize the beam waist within a 10 MHz bandwidth germanium 
acousto-optic modulator (AOM). The first diffracted (modulated) beam was recollimated (L2) and passed through
a 1~mm aperture, to reject the main beam and higher order diffracted beams, and focused (L3) onto the \paperCone{} sample held in superfluid helium. A portion of the resulting 1s:T$_2$:$\Gamma_7$ $\Rightarrow$ 1s:A luminescence signal was captured by an elliptical mirror and sent through 2440~nm and 2850~nm long-pass filters, Filter 1 and Filter 2, to selectively pass 1s:T$_2$:$\Gamma_7$ $\Rightarrow$ 1s:A light into an MCT detector. A lock-in measurement was applied to the detected signal using the AOM driving frequency as the reference.}
	\label{fig:setup}
\end{figure}

To directly measure the excited state lifetime, we performed a modulated excitation experiment~\cite{Mallawaarachchi1987} using a continuous-wave, single-frequency laser modulated by an acousto-optic modulator (AOM). The measurement configuration is shown in Fig.~\ref{fig:setup}. The laser was brought into resonance with the 1s:A $\Rightarrow$ 2p$_\pm$ transition (4578 cm$^{-1}$, or 2184~nm), as in Sec.~\ref{sec:pho}, to efficiently pump to 1s:T$_2$:$\Gamma_7$ via the 2p$_\pm$ state. This pump laser was sinusoidally modulated with a germanium AOM (IntraAction AGM-802A9) with a nominal bandwidth of 10 MHz, which was increased to 20 MHz by reducing the laser spot size using a converging lens pair. Approximately
400 mW of laser light was incident on the sample.
The resulting 1s:T$_2$:$\Gamma_7$ $\Rightarrow$ 1s:A luminescence from the sample \paperCone{} (see Supplementary Materials~\cite{DPMSuppMat}) was spectrally filtered and detected using an MCT detector (VIGO Systems, PVI-4TE-1-0.5x0.5), and fed into a lock-in with the AOM modulation drive as its reference.

\begin{figure}[h]
	\centering
	\adjincludegraphics[width = 1\columnwidth, clip]{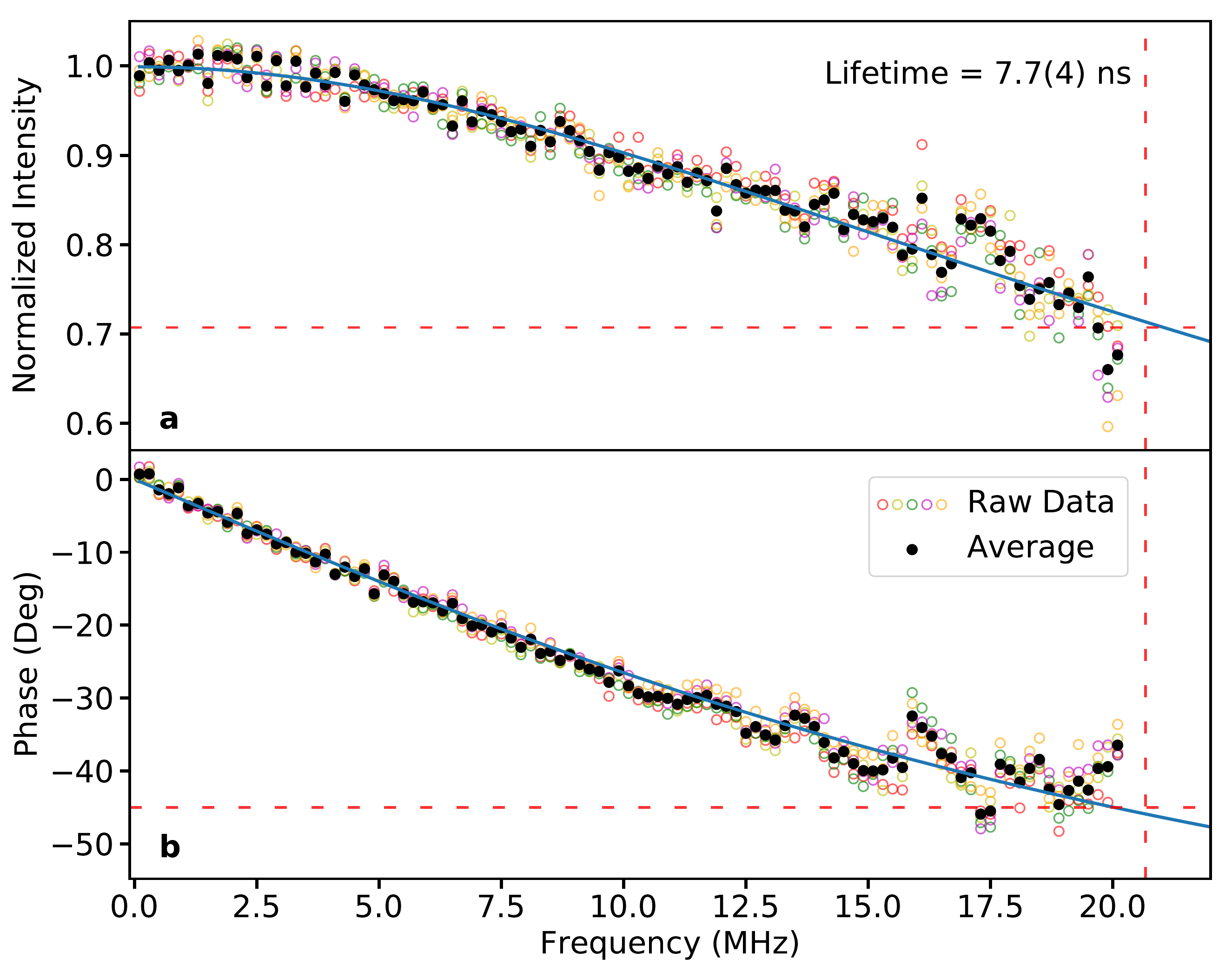}
	\caption{Excited state lifetime measurement of the 1s:T$_2$:$\Gamma_7$ state of Se$^+$ detected through modulation-frequency-dependent luminescence. Five separate datasets are plotted in open circles and their average is indicated by closed black circles.
	(\textbf{a}) Amplitude response of the photoluminescence as a function of excitation modulation frequency.  The averaged amplitude data is fit with Eqn.~\ref{eq:amplife} (blue curve) and the data is normalized to the fit amplitude. (\textbf{b}) Phase response of the photoluminescence as a function of excitation modulation frequency. The phase lag between the AOM drive signal and the luminescence signal is fit with Eqn.~\ref{eq:phaselife} (blue curve). Red dashed lines intersect the data at the critical modulation frequencies, at which the normalized fit amplitude has dropped to 1/$\sqrt{2}$ and phase has lagged by 45$^{\circ}$, revealing a T$_1$ time for the 1s:T$_2$:$\Gamma_7$ excited state of \paperexlifetime{}~ns.}
	\label{fig:rad}
\end{figure}

After correcting for the instrumental frequency response by measuring scattered pump laser light,
the frequency dependence of the resulting signal revealed the excited state lifetime.
At frequencies much lower than the inverse of the excited state lifetime the system has time to equilibrate and a high AC photoluminescence signal is detected, whereas at higher frequencies the AC signal amplitude will drop. Alternatively put, the system behaves as a low-pass filter with a characteristic amplitude ($A$) and phase ($\Theta$) response, as a function of modulation frequency, $f$, given by~\cite{Mallawaarachchi1987}:
\begin{align}
A &= \frac{1}{[1+(2\pi f\, \rm{T}_1)^2]^{1/2}},
\label{eq:amplife} \\
\Theta &= -\tan^{-1}\left(2\pi f\, \rm{T}_1\right)
\label{eq:phaselife}
\end{align}
where T$_1$ is the decay time of the optically excited state. The resulting 
data, corrected for the system response,
are shown in Fig.~\ref{fig:rad}. The characteristic amplitude and phase drop-off points, at $1/\sqrt{2}$ and $45^{\circ}$, agree and reveal a T$_1$ time for the 1s:T$_2$:$\Gamma_7$ excited state to be \paperexlifetime{}~ns. 

This gives a radiative efficiency of \paperradeff{}$\%$ when compared to the radiative lifetime of \papertautotal{}~\us{}, as well as a homogeneous linewidth of $0.00069(4)$~cm$^{-1}$. 
However, as thermally activated transitions to higher 
excited states are possible~\cite{Morse2017} this homogeneous linewidth
is likely to be a lower bound. 
For the purposes of estimating coupling cooperativity~\cite{DPMSuppMat}
between 
the Se$^+$ 1s:A~$\Leftrightarrow$~1s:T$_2$:$\Gamma_7$ transition
and a photonic cavity 
we use the upper bound
determined by hole burning, 0.001~cm$^{-1}$. 
With a ZPL dipole moment of $\mu$ = \paperTDM{} Debye, 
a Se$^+$ spin in the mode maximum of a cavity with an unloaded Q-factor of 1.5~$\times$~10$^4$ and a modal volume, 
$V~=~(\lambda/n)^3$,
would display a cooperativity of C~=~1.

\subsection{Raman spectroscopy}
\label{sec:raman}

The 1s:A $\Leftrightarrow$ 1s:T$_2$:$\Gamma_7$ transition amounts to at least a seven phonon transition, and yet results from Sec. \ref{sec:lifetime} show that relaxation from 1s:T$_2$:$\Gamma_7$ is predominantly nonradiative. Although Altarelli \cite{Altarelli1983} predicted the Se$^+$ 1s:E state to lie above 1s:T$_2$, the Se$^+$ 1s:E state has not yet been experimentally observed. It is conceivable, however highly unusual, that 1s:E lies below 1s:T$_2$. 
If 1s:E were to lie below 1s:T$_2$:$\Gamma_7$ it could provide a nonradiative decay pathway which could
account for the low radiative efficiency of the 1s:T$_2$:$\Gamma_7$~$\Leftrightarrow$~1s:A transition.

The 1s:A $\Leftrightarrow$ 1s:E transition is both EMT and symmetry-forbidden, in contrast with 1s:A $\Leftrightarrow$ 1s:T$_2$:$\Gamma_7$ which is symmetry-allowed, and so indirect methods are needed to deduce the binding energy of the 1s:E state of both Se$^0$ and Se$^+$. In the neutral charge state, Se$^0$, the location of the 1s:E state has been shown to lie above 1s:T$_2$,
which for the neutral state Se$^0$ splits into levels 1s:$^1$T$_2$ and 1s:$^3$T$_2$ (see Ref.~\cite{Grossmann1987}).
The position of 1s:E was extrapolated from strain-induced hybridization of the 1s:E and 1s:$^1$T$_2$ levels~\cite{Grossmann1987}, with a projected unstrained binding energy of 31.4 meV, corresponding to a 1s:A $\Leftrightarrow$ 1s:E transition of 2220~cm$^{-1}$. Here we show the results of Raman spectroscopy in an effort to observe forbidden transitions in both Se$^+$ and 
Se$^0$, specifically the 1s:A $\Leftrightarrow$ 1s:E transition which
has been observed for shallow donors~\cite{Wright1967}.

\begin{figure}[h]
\centering
\adjincludegraphics[width = 1\columnwidth, clip]{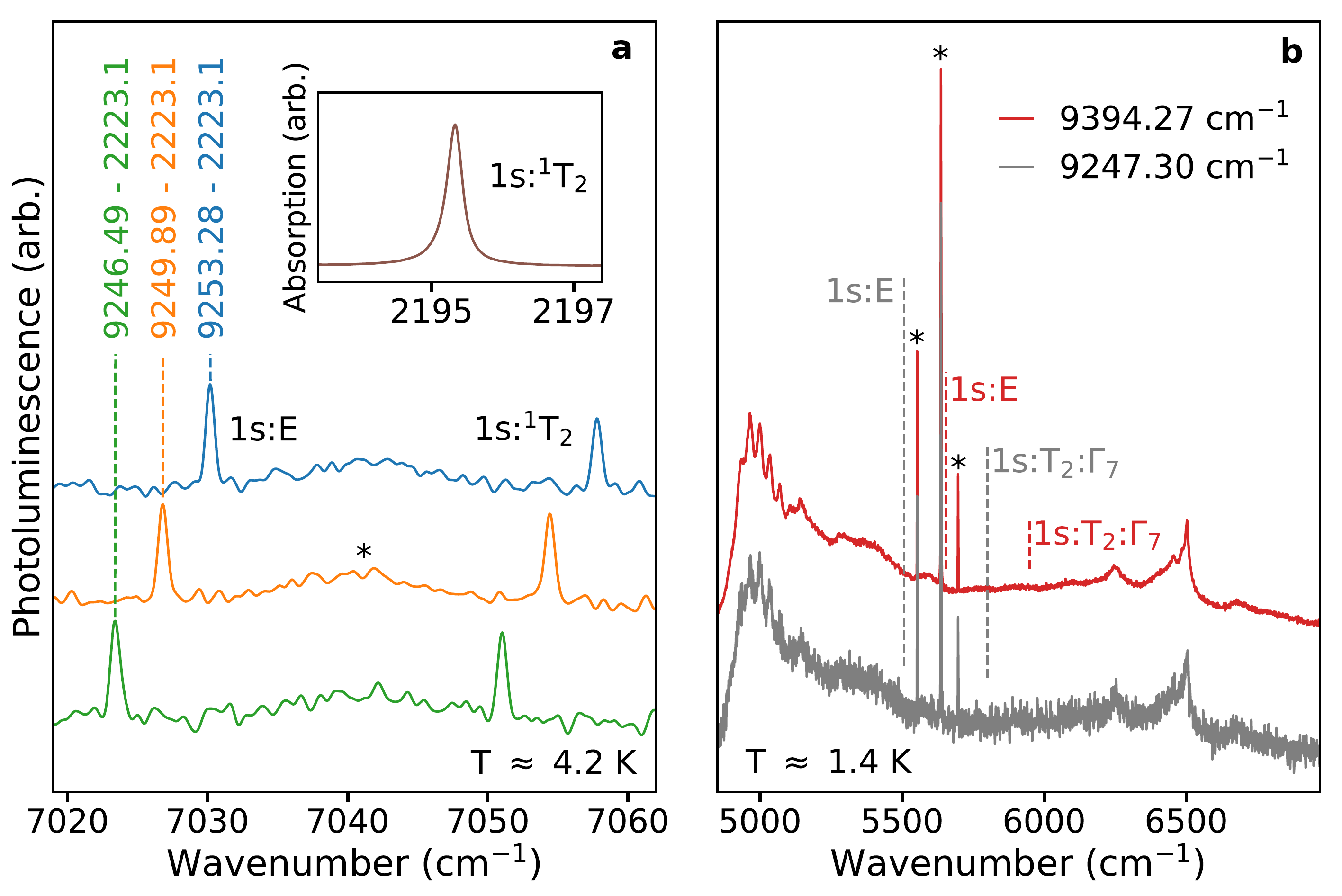}
\caption{(\textbf{a}) Raman spectroscopy of the Se$^0$ region showing features whose shifts in energy match the shifts in tunable laser energy. (Applied laser energy from bottom to top: 9246.49, 9249.89, and 9253.28$~\rm{cm}^{-1}$). The left (right) peaks correspond to Raman scattering from the 1s:A $\Leftrightarrow$ 1s:E (1s:A $\Leftrightarrow$ 1s:$^1$T$_2$) transition with an average measured offset of 2223.1$~\rm{cm}^{-1}$ (2195.5$~\rm{cm}^{-1}$) from the applied laser energy. A broad photoluminescence feature of unknown origin is labeled by an $\ast$. Spectra were collected with a resolution of 0.5 cm$^{-1}$ using a liquid nitrogen-cooled Ge diode detector. (\textbf{Inset}) The offset due to 1s:A $\Leftrightarrow$ 1s:$^{1}$T$_2$ matches the observed energy of the 1s:A $\Leftrightarrow$ 1s:$^{1}$T$_2$ transition seen in absorption.
(\textbf{b}) Raman spectroscopy of the Se$^+$ donor region revealing no signs of 1s:E or any other Raman-shifted transitions. Photoluminescence lines corresponding to bound exciton features of unknown origin are marked with an $\ast$. Expected Raman feature positions are indicated with dashed lines assuming the 
theoretically predicted binding energy of 1s:E~\citep{Altarelli1983} and the known transition energy of 1s:A $\Leftrightarrow$ 1s:T$_2$:$\Gamma_7$. Spectra were collected with a resolution of 1.0 cm$^{-1}$ using a liquid nitrogen-cooled InSb detector with a cryogenically mounted band-pass filter.
}
\label{fig:raman}
\end{figure}

Raman spectra of the \paperEone{} sample~\cite{DPMSuppMat} were measured using a Bruker IFS 125HR FTIR spectrometer using tunable narrowband 1080~nm ($\sim$9260 cm$^{-1}$) and 1064 nm ($\sim$9400~cm$^{-1}$) excitation sources, amplified using an IPG Photonics amplifier (YAR-10K-1064-LP-SF), a CaF$_2$ beam splitter, and detected using either a liquid nitrogen-cooled Ge diode detector (for Se$^0$ Raman experiments) or a liquid nitrogen-cooled InSb detector (for Se$^+$ Raman experiments) with a band-pass filter mounted in the InSb detector's cryogenic assembly to reduce incident room temperature blackbody radiation and increase sensitivity (although the 
cold-filtered
InSb was still much less sensitive than the Ge diode detector).
In the detection arm, 1150 and 1200~nm long pass filters were used for laser rejection, with an additional 1100~nm long pass filter used in the Se$^0$ Raman experiments. 

In Fig.~\ref{fig:raman}a we see the results of Raman spectroscopy centred near ($9260-2220$) cm$^{-1}$ where we expect to observe Raman
features corresponding to the 1s:A~$\Leftrightarrow$~1s:E transition of Se$^0$ when driving with laser light near 1080~nm. 
We observe a feature which shifts linearly with laser frequency closely matching the projected value for the 1s:A~$\Leftrightarrow$~1s:E transition. Although unexpected from shallow donor Raman measurements,
we also observe a Raman-active feature that
matches the measured value of the 1s:A~$\Leftrightarrow$~1s:$^{1}$T$_2$ transition.

We measure an average shift from the laser position of 2223.1(5)~cm$^{-1}$ corresponding to 1s:A $\Leftrightarrow$ 1s:E which agrees with the projected strain-free transition frequency of 2220~cm$^{-1}$ from Ref.~\cite{Grossmann1987}. We measure an average shift from the laser position of 2195.5(5)~cm$^{-1}$ which agrees with the 1s:A $\Leftrightarrow$ 1s:$^{1}$T$_2$ transition energy of 2195.4(5)~cm$^{-1}$ directly observed in absorption (See inset of Fig.~\ref{fig:raman}a). 

In Fig.~\ref{fig:raman}b we show the spectral region where one would expect to observe Raman transitions associated with the 1s:A $\Leftrightarrow$ 1s:E transitions of Se$^+$. Energies labelled 1s:E, denoted by dashed vertical lines in Fig.~\ref{fig:raman}b, are based on the calculations of Altarelli \cite{Altarelli1983} who predicted the 1s:E level of Se$^+$ to have a binding energy of $\sim$130~meV, corresponding to a 1s:A $\Leftrightarrow$ 1s:E transition near 3740 cm$^{-1}$. We note no observable feature shifts over the broad range we would expect to detect Raman Se$^+$ transitions. It is possible that 1s:E is simply very broad making it extremely difficult to observe. The 1s:A $\Leftrightarrow$ 
1s:T$_2$:$\Gamma_7$ transition was not observed, which agrees with similar shallow donor Raman experiments.
The precise binding energy of 1s:E level of Se$^+$ remains the subject of future investigation. 

\section{Conclusion}
We have demonstrated that a variety of performance metrics of the $^{77}$Se$^+$ spin-photon interface, built upon its 1s:A $\Leftrightarrow$ 1s:T$_2$:$\Gamma_7$ transition, are more favourable than previously thought. A number of key properties of this interface were examined and shown to have encouraging features, including long spin T$_1$ lifetimes exceeding 4.6(1.5) hours at low temperatures and near Earth's magnetic field, a larger transition dipole moment of \paperTDM{}~Debye, a 1s:T$_2$:$\Gamma_7$ excited state lifetime of~\paperexlifetime{}~ns, a total radiative efficiency of \paperradeff{}~\%, and a zero phonon line fraction of \paperzpleff{}~\%. These results imply that the spin-dependent cavity cooperativity threshold of 1 may be crossed 
with routinely achievable photonic cavities having mode volumes of $\sim\!(\lambda/n)^3$ and Q-factors of 1.5~$\times$~$10^4$.
A broad variety of silicon quantum technologies may be built based upon this key and highly sought-after spin-dependent nonlinearity. 
\nocite{Devyatykh2008,Kim1979,Schibli1967,Kim1979n2,pichler,Zakel2011,Yu1988,Sennikov2005,Hilborn1982,Frey2006}

\subsection*{Acknowledgments}
The $^{28}$Si samples used in this study were prepared from Avo28 crystal produced by the International Avogadro Coordination (IAC) Project (2004-2011) in cooperation among the BIPM, the INRIM (Italy), the IRMM (EU), the NMIA (Australia), the NMIJ (Japan), the NPL (UK), and the PTB (Germany). We thank Eundeok Mun for sealing the quartz ampules used to make the samples. We also thank Dev Sharma for synthesizing
the selenium dioxide used to make the samples.

\subsection*{Funding}

This work was supported by the Natural Sciences and Engineering Research Council of Canada, the Canada Research Chairs program, the Canada Foundation for Innovation and the British Columbia Knowledge Development Fund.

%


\end{document}